# Artificial-Intelligence-Based Design (AI-D) for Circuit Parameters of Power Converters

Xinze Li, *Student Member*, *IEEE*, Xin Zhang, *Senior Member*, *IEEE*, Fanfan Lin, *Student Member*, *IEEE*, Frede Blaabjerg, *Fellow*, *IEEE*

*Abstract* --- Parameter design is significant in ensuring a satisfactory holistic performance of power converters. Generally, circuit parameter design for power converters consists of two processes: analysis and deduction process and optimization process. The existing approaches for parameter design consist of two types: traditional approach, computer-aided optimization (CAO) approach. In the traditional approaches, heavy human-dependence is required. Even though the emerging CAO approaches automate the optimization process, they still require manual analysis and deduction process. To mitigate human-dependence for the sake of high accuracy and easy implementation, an artificial-intelligence-based design (AI-D) approach is proposed in this paper for the parameter design of power converters. In the proposed AI-D approach, to achieve automation in the analysis and deduction process, simulation tools and batch-normalization neural network (BN-NN) are adopted to build data-driven models for the optimization objectives and design constraints. Besides, to achieve automation in the optimization process, genetic algorithm is used to search for optimal design results. The proposed AI-D approach is validated in the circuit parameter design of the synchronous Buck converter in the 48 V to 12 V accessory-load power supply system in electric vehicle. The design case of an efficiency-optimal synchronous Buck converter with constraints in volume, voltage ripple and current ripple is provided. In the end of this paper, feasibility and accuracy of the proposed AI-D approach have been validated by hardware experiments.

## I. INTRODUCTION

Power converters have been increasingly applied nowadays in both industries and our daily life. Power converters can regulate power transmission and alter the form of voltage and current [1]. In industrial applications, power converters such as DC-DC converters and inverters are the critical enablers in renewable energy systems [2], wireless power transfer [3] and DC microgrids [4]. Even in our daily life, the applications of power converters are omnipresent, such as electric vehicle [5], solar PV [6], etc.

Manuscript received February 03, 2021; revised March 24, 2021; revised May 17, 2021; accepted May 26, 2021. This work was supported by Start-up grant of Professor Zhang at Zhejiang University. (*Corresponding author: Xin Zhang*).

Xinze Li is with the School of Electrical and Electronic Engineering, Nanyang Technological University, Singapore 639798, Singapore. (e-mail: xinze001@e.ntu.edu.sg).
Xin Zhang is with the College of Electrical Engineering, Zhejiang University, Hangzhou 310027, China, and with Hangzhou Global Scientific and Technological Innovation Center, Zhejiang University, Hangzhou 310058, China. (e-mail: zhangxin_ieee@163.com).
Fanfan Lin is with ERI@N, Interdisciplinary Graduate Program, Nanyang Technological University, Singapore 639798, Singapore. (e-mail: fanfan001@e.ntu.edu.sg).
Frede Blaabjerg is with the Department of Energy Technology, Aalborg University, Aalborg, Denmark DK-9220, Denmark. (e-mail: fbl@et.aau.dk).

To ensure the satisfactory performance of power converters in all the applications, the circuit parameters of power converters should be carefully designed. And efficiency, size, cost, reliability, ripples and transient response are several commonly-adopted design objectives [7]–[10]. To achieve a comprehensively better performance, usually multiple design objectives are considered simultaneously, which renders the parameter design for power converters significant and challenging.

Generally, the parameter design for power converters consists of two processes: analysis and deduction process, and optimization process. Up to now, there have been two main approaches in the parameter design of power converters, which are traditional human-dependent design approach [11], [12] and computer-aided optimization (CAO) design approach [8], [13], [14]. For traditional human-dependent design approach, the analysis of optimization objectives and design constraints are conducted by engineers totally to deduce mathematical expressions [11]. And the optimization process will also be conducted with repetitious manual trials and errors. The major drawbacks, from which the traditional human-dependent design approach is suffering, are heavy work burden and the low accuracy due to approximation during the analysis and deduction process [11]. As for the CAO approach which has been proposed in the past two decades, the optimization process has been carried out with some optimization algorithms on computers, such as the recent popular particle swarm optimization algorithm, genetic algorithm and ant colony algorithm [14]. However, the analysis and deduction process for optimization objectives and design constraints are still highly human-dependent, resulting in low accuracy and large amount of consumed time [13].

To relieve the manpower burden in the analysis and deduction process of parameter design, neural network (NN), which is an artificial-intelligence (AI) technique, is an ideal technique to be adopted. Emulating the adaptive connections of neurons in brain, NN has the capability to learn from and interpret external data by tuning its adjustable weights. The easily scalable structure of NN makes it possible to learn any complicated non-linear functions with any accuracy. Due to these advantages, NN has been widely adopted in motor optimization [15], modulation strategies [16], and control [17], [18]. For instance, [15] adopts extreme learning machine to fit the results of finite element analysis to optimize motor. [16] utilizes NN to realize optimal selective harmonic elimination. NN serves as the model predictive controller for modular multilevel converters in [18].

However, NNs utilized in the current literature of power converters are mostly basic networks, the generalization accuracy of which in unseen data still has room for improvements. To realize a higher prediction accuracy in unseen data, the structure of NN should be improved.

As inspired by the strong learning capability of NN, to deal with the above difficulties which the existing parameter design approaches for power converters are facing, an artificial-intelligence-based design (AI-D) approach is investigated in this paper. In the proposed AI-D approach, automation in the two procedures in parameter designs can be achieved: automation in the analysis and deduction process and automation in the optimization process. The analysis and deduction process will be conducted by data-driven models with the assistance of simulation tools and batch-normalization neural network (BN-NN). And the evolutionary algorithm (EA) will take the responsibility for the optimization process. With BN-NN and the automatic fashion in these two procedures, the parameter design for power converters can achieve accurately optimal design results as well as a large extent of freedom for engineers.

The rest of this paper is organized as follows. The problems in the existing parameter design approaches for power converters and the proposed solutions are described in Section II. The detailed process of the proposed AI-based design approach is elaborated in Section III. A design case is illustrated in Section IV and experimental results are presented in Section V. And Section VI draws the conclusion for this paper.

## II. PROBLEM DESCRIPTIONS FOR THE PARAMETER DESIGN APPROACHES FOR POWER CONVERTERS AND THE PROPOSED SOLUTIONS

### A. Problems in the Existing Circuit Parameter Design Approaches for Power Converters

In the design of circuit parameters for power converters, after specifying the design objectives and operating conditions, there are two major important procedures. The first one is the analysis and deduction process for the optimization objectives and design constraints, after which the mathematical expressions for optimization objectives and design constraints will be obtained. And the second one is the optimization process to find out the circuit parameters which can realize optimal performance in the optimization objective without breaking the design constraints.

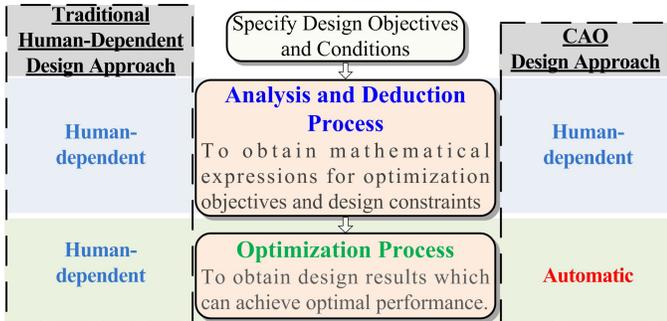

Fig. 1. Realization of design process (human-dependence or automation) in two kinds of parameter design approaches for power converters: traditional human-dependent design approach and CAO design approach.

In the traditional human-dependent design approach, as described with the left side in Fig. 1, analysis and deduction process and optimization process are both carried out by engineers. To derive the mathematical expressions for the targeted optimization objective and design constraints, many approximations will be taken for the sake of analytical convenience. For example, to derive the operating point of two-switch forward converter in [11], the switches are assumed to be ideal. Besides, as introduced by Lin et. al., in the deduction of output impedance of dual active bridge converter, only $0^{th}$ and $1^{st}$ order terms in Fourier expansion are considered [19]. Even though sometimes approximations are removed for the sake of high accuracy, the computational burden and complexity are heavy for engineers [12], [20]. When it comes to the optimization process, repetitious manual trials and errors are also time-consuming and have no guarantee for optimization accuracy [21].

Time has seen great improvements since the development of computer science has contributed to the CAO parameter design approach in the last two decades, which is presented in the right side of Fig. 1. The CAO design approach has realized automation in the optimization process with the assistance of optimization algorithms [8], [13], [22]–[24]. For instance, [8] modifies the archived multi-objective simulated annealing algorithm (AMOSA) to achieve multi-objective design of Buck converter. [22] adopts Monte-Carlo for the optimal DC-DC converter in automobiles. Liu et. al. deduces the capacitor lifetime through the analysis of its internal temperature rise, analyzes the volume of $L$ and $C$ based on the inductance and capacitance definitions and their geometric features, and adopts NSGA-II algorithm for a good comprehensive parameter design [25].

Even though CAO design approach has freed engineers from repetitious trials and errors to achieve optimal design results, the human-dependence in the analysis and deduction process remains unsolved. To obtain the mathematical expressions of the optimization objective and design constraints, complicated and time-consuming analysis cannot be avoided, which is, at the same time, suffering from accuracy concern. Besides, the nature of optimization algorithms adopted in CAO is intrinsically a set of rigid instructions to find optimal solutions for a given function, so CAO approaches have no capability to learn from and interpret external data to grow intelligence. Therefore, CAO approaches cannot be considered directly as AI.

### B. The Proposed Solutions for the Automated Design for the Circuit Parameters of Power Converters

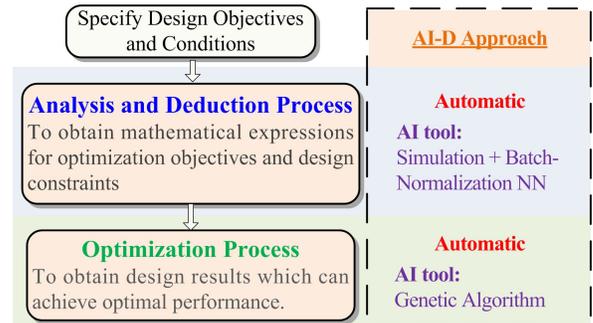

Fig. 2. Realization of design process in the proposed AI-D approach for the circuit parameters of power converters.

To deal with the problems including heavy work burden and low accuracy in the parameter design for power converters, which is attributable to the high level of human-dependence, an artificial-intelligence-based design (AI-D) approach is specially put forward. As shown in Fig. 2, this design approach can carry out both the analysis and deduction process and the optimization process in an automatic fashion, facilitating an easy-implemented and accurate design.

(*a*) *Automation in Analysis and Deduction Process*

To eliminate human-dependence in analysis and deduction process, simulation and batch-normalization NN (BN-NN) are adopted to build a data-driven model, as described with Fig. 3.

Simulation models will be built to evaluate the performance in the optimization objectives as well as the performance in design constraints under different selections of design parameters. With simulation models, a neural network will be provided with data set for training. After training with limited performance data, NN can serve as an accurate data-driven model because it can figure out any complicated and nonlinear relationships between design parameters and performance. Thus, it can be used to predict the performance of new designs.

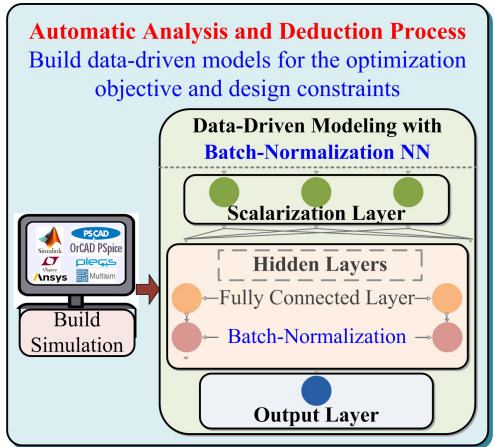

Fig. 3. Automation in the analysis and deduction process in the proposed AI-D.

For the sake of better generalization accuracy in predicting unseen design performance, BN-NN is specially adopted for building the data-driven models. It consists of three kinds of layers: scalarization layer, hidden layers, and output layer. BN-NN is named because of its special batch-normalization layer in the hidden layer. The batch-normalization layer aims to avoid the over-fitting problem, which will seriously deteriorate the NN accuracy. Compared with other techniques in avoiding over-fitting, such as L2 weight-decay regularization [26] and dropout [27], batch-normalization layer [28] does not introduce extra hyperparameters, and thus is simpler for the tuning of NN. Basically, the batch-normalization layer applies normalization to layer inputs, and it can adaptively tune the degree of normalization, with which the over-fitting problem is largely mitigated [28].

With simulation models and BN-NN, the analysis and deduction process in the design process can be conducted in an automatic fashion, freeing engineers from heavy work burden. And more accurate performance evaluation can be guaranteed compared to the traditional manual analysis process with approximation. The special structure of BN-NN also guarantees better generalization accuracy in predicting unseen design performance.

(*b*) *Automation in Optimization Process*

Apart from automation in analysis and deduction process, optimization process in which design results are finalized to achieve optimal performance can also be carried out automatically.

Genetic algorithm (GA), one of the popular evolutionary algorithms, is chosen to take the responsibility of parameter optimization. This choice is considered from the special characteristic of parameter design problem for power converters. When a power converter is designed, the design parameters may lie in continuous space or discrete space. For example, when switching frequency is designed, usually continuous design space is considered. Whereas when inductors or capacitors are designed, only discontinuous and unconnected values can be chosen because of practical components, so they lie in discrete design space. The mix of design parameters in both discrete and continuous space contributes to a mixed-integer optimization problem. Among popular EA (PSO, GA and ACO), PSO is suitable for continuous optimization [29], and ACO aims at discrete optimization [30]. Fortunately, GA performs the best in mixed-integer optimization and it also enjoys fast convergence speed [23]. Thus, GA is suitable for AI-D approach for the parameter design of power converters.

With the adopted GA, global optimal design parameters will be located to achieve good performance in the targeted optimization objective without breaking design constraints. Also, its fast convergence speed is helpful for a prompt design of power converters.

In a word, as discussed above, the proposed AI-D approach is able to realize a high level of automation in two procedures: analysis and deduction process and optimization process. The special BN-NN adopted in AI-D ensures high predicting accuracy in unseen design parameters. Hence, the proposed AI-D approach allows for high accuracy and easy-implementation in the parameter design of power converters.

III. AI-D APPROACH FOR THE PARAMETER DESIGN OF POWER CONVERTERS

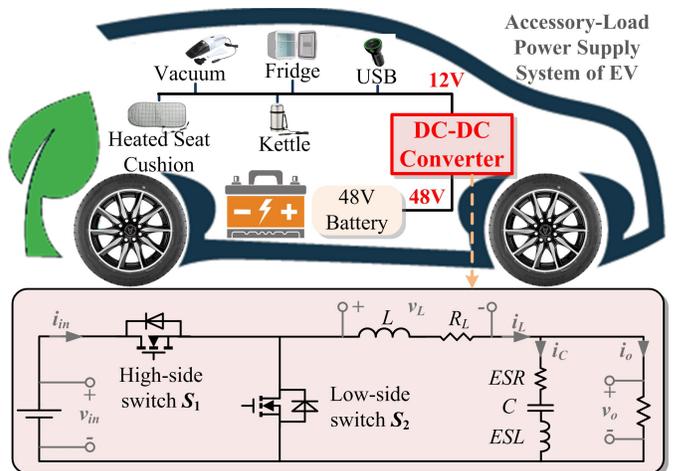

Fig. 4. The parameter design of synchronous Buck converter in the 48 V to 12 V

accessory-load power supply system in EV via the proposed AI-D approach.

The proposed AI-D methodology can be applied to design circuit parameters for all kinds of power converters with no limitations on the application backgrounds. In this paper, for the sake of easy understanding, the proposed AI-D approach is validated in the circuit parameter design of the synchronous Buck converter in the 48 V to 12 V accessory-load power supply system in electric vehicle (EV) [31].

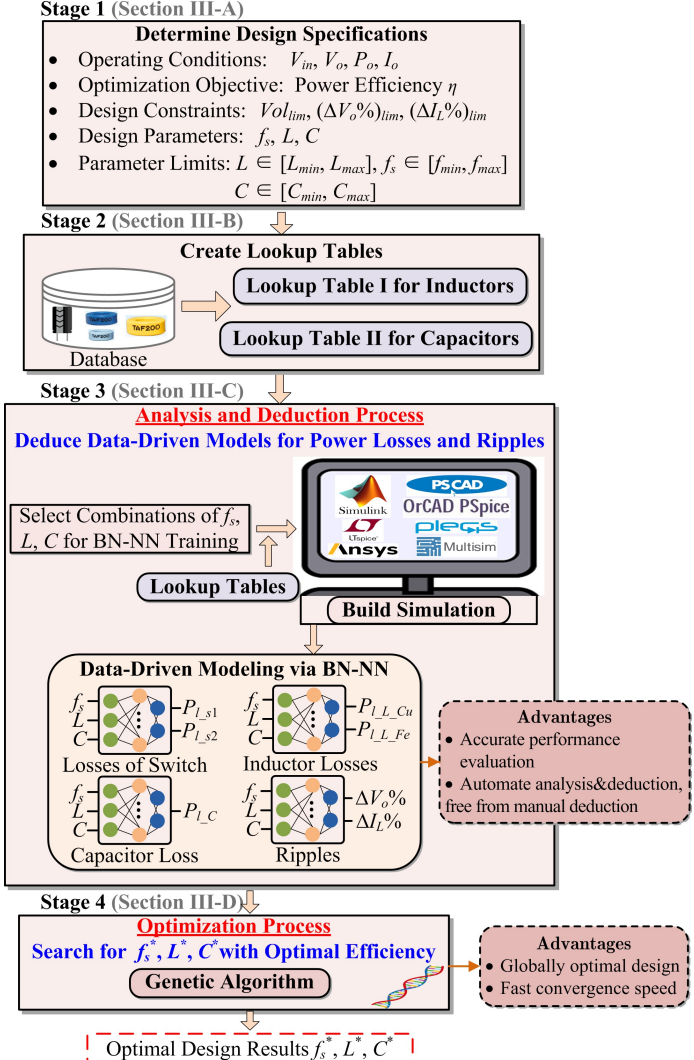

Fig. 5. Flowchart of the proposed AI-D approach applied in the parameter design of an efficiency-optimal synchronous Buck converter in EV.

In EV, the 48 V to 12 V power supply system is increasingly adopted to realize the power conversion from the provided 48 V power supply to the required 12 V accessory-load voltage [32], as shown in Fig. 4. To facilitate this power conversion process, usually a synchronous Buck topology is adopted thanks to its good efficiency and simple structure [31], [32]. When it is designed, efficiency is one important performance indicator to ensure high power transfer efficiency. In addition, volume and ripples are considered for the sake of a compact and reliable design to ensure good holistic performance. In this paper, the design process takes efficiency as the optimization objective and takes volume and ripples as design constraints as an example. It should be noted that if any other objectives or constraints, like cost or reliability, are preferred in other design backgrounds, this proposed AI-D approach can still be applicable and only some customizations are needed.

In this section, the AI-D approach to design the circuit parameters for the efficiency-oriented synchronous Buck converter in EV is elaborated, which considers volume and ripple as design constraints. This design process contains 4 stages, as shown in Fig. 5.

### A. Stage 1: *Determine Design Specifications*

Before further design process, determining all the design specifications is the first stage.

The design conditions should be firstly figured out, including input voltage $V_{in}$, output voltage $V_o$, output power $P_o$ and output current $I_o$. Power efficiency is the optimization objective, and the design constraints include volume constraint $Vol_{lim}$, voltage ripple constraint $(\Delta V_o\%)_{lim}$ and current ripple constraint $(\Delta I_L\%)_{lim}$. The parameters that need to be designed contain switching frequency $f_s$, inductance $L$ and capacitance $C$. The limits of design parameters are also necessary as $[f_{min}, f_{max}]$, $[L_{min}, L_{max}]$ and $[C_{min}, C_{max}]$.

The volume as a design constraint only takes the size of the inductor and capacitor ($Vol_L$, $Vol_C$) into considerations because they will be greatly influenced by the choices of design parameters [8]. The minor effects of parameter design on the volume of other parts, like PCB board, are neglected.

### B. Stage 2: *Create Lookup Tables for Inductors and Capacitors*

In Stage 2, since the values of the inductor and capacitor are both design parameters, lookup tables have to be created for them. With the considerations of practical components, lookup tables store all the reachable values of inductors and capacitors and aim to bridge the gap between theoretical analysis and reality.

For inductors, Lookup Table I maps the value of inductor $L$ to the geometric and magnetic features of the selected core. For capacitors, Lookup Table II maps the value of capacitor $C$ to the features of the selected capacitor components.

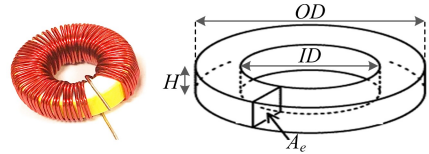

Fig. 6. Toroidal inductor applied in Buck converter.

The core with toroidal shape, as shown in Fig. 6, is chosen as an example. The process to create Lookup Table I for inductors is described with Fig. 7. After the specification of the fill factor of inductor core [33], $K_u$, with the geometric and magnetic features of cores obtained from core database, the maximum number of turns $N_{max\_core}$ and the maximum inductance $L_{max\_core}$ each core can reach are computed with (1) and (2), respectively. In (1) and (2), core inner diameter $ID$, nominal inductance $A_L$ are obtained with database, and $A_W$ is the area of wire. Afterwards, $L_{max\_core}$ is sorted in an ascending order. $L_{max\_core}$ partitions the selection of inductor core. For instance, if $L$ is lower than or equal to $L_{max\_core1}$, for the sake of smaller size of the core, core 1 is selected; If $L$ is greater than $L_{max\_core1}$, to avoid core saturation, core 2 is selected. Subsequently, a lookup table on the choices of cores for the reachable values of inductors can be created.

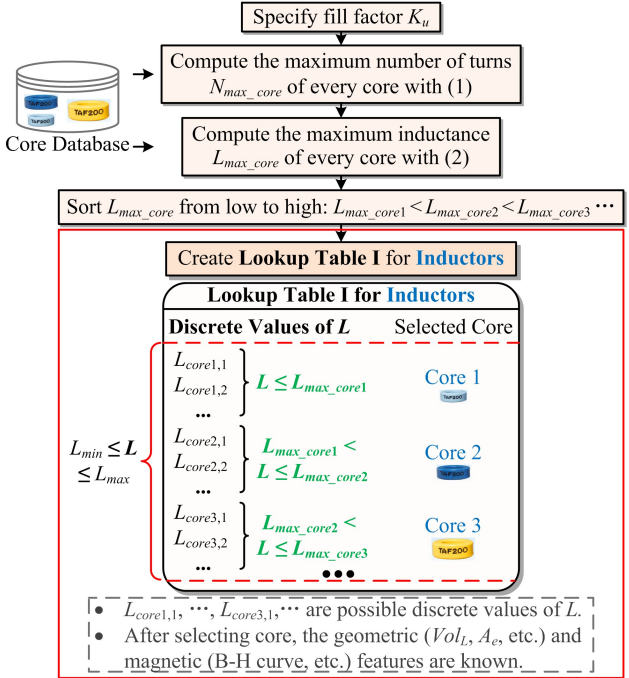

Fig. 7. Create Lookup Table I for inductors.

$$N \leq K_u \cdot \frac{\pi ID^2}{4A_W} = N_{max\_core} \quad (1)$$

$$L_{max\_core} = \frac{A_L}{1000} N_{max\_core}^2 \ \mu H \quad (2)$$

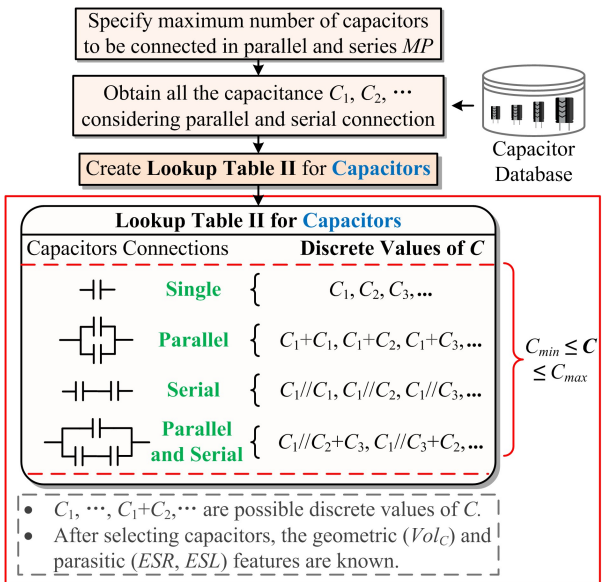

Fig. 8. Create Lookup Table II for capacitors.

When it comes to the creation of a look up table for capacitors, flowchart in Fig. 8 is followed. To realize more values of capacitors with the limited practical components, possible parallel and serial connections of single capacitors are considered. After the specification of maximum number of components for parallel and serial connections $MP$, all available values of capacitors can be obtained. And then Lookup Table II on the different ways of connections for different values of capacitors can be built. If there are different combinations for the same designed value (e.g., for 660 $\mu$F, there can be 220 $\mu$F * 3 and 330 $\mu$F * 2), the combination with smallest volume is selected.

In this design case, for the sake of illustration convenience, only toroidal cores and electrolytic capacitors are taken as design examples. If more types of cores and capacitors are considered, only minor adjustments are required: one lookup table will be created for one type of cores or capacitors, so different types of cores and capacitors will have different lookup tables. With the created different lookup tables which store the information of different types of cores and capacitors, the types of cores and capacitors can be incorporated as the design parameters to be optimized in the process of AI-D.

### C. Stage 3: Build Data-Driven Models for Power Losses, Voltage Ripple and Current Ripple

In Stage 3 of the proposed AI-D approach, data-driven models of power losses, voltage and current ripples are automatically built with simulations and BN-NN. In this procedure, simulation, which can provide accurate performance evaluation, evaluates the power loss and ripple performance of different designs. After the training of BN-NN with these simulation results, BN-NN will serve as accurate data-driven models for power losses and ripples. The detailed flowchart of Stage 3 is provided in Fig. 9 which contains 3 steps.

(a) *Step* 1: *Select Combinations of $f_s$, L, C for BN-NN Training*

In Step 1 of Stage 3, combinations of design parameters $f_s$, L, C should be firstly selected. $f_s$, L and C are uniformly selected within $[f_{min}, f_{max}]$, $[L_{min}, L_{max}]$ and $[C_{min}, C_{max}]$ for $N_1$, $N_2$ and $N_3$ number of points, respectively. As a result, the total number of combinations of $f_s$, L, C generated is $N_1 \times N_2 \times N_3$.

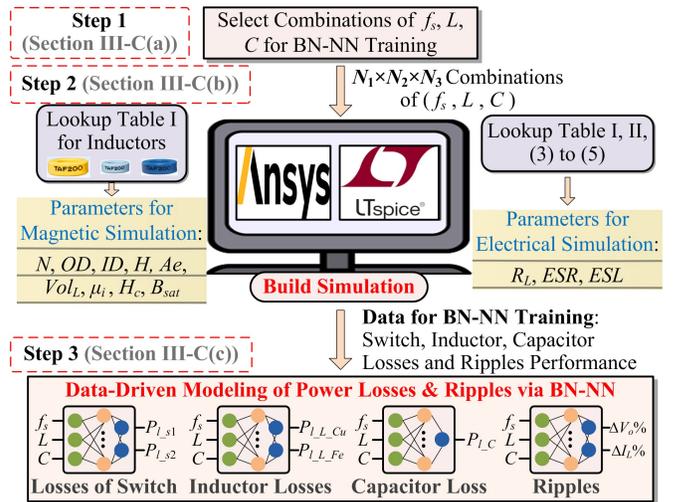

Fig. 9. Detailed realization of Stage 3: build data-driven models for power losses and ripples with simulations and BN-NN.

(b) *Step* 2: *Build Simulation for Power Losses and Ripples of Selected Combinations to Generate Data for BN-NN Training*

Step 2 of Stage 3 aims at building simulations for performance evaluation of the selected combinations of design parameters in Step 1 to generate data for training BN-NN.

The performance indicators that need evaluations include power losses, voltage ripple, current ripple and volume. Among them, power losses will be analyzed by magnetic simulation and electrical simulation. Voltage ripple and current ripple will be evaluated by electrical simulation. And volume will be obtained from the created lookup tables.

To build the magnetic simulation, the features of inductor core (number of turns $N$, core geometry $OD$, $ID$, $H$, $Ae$, inductor volume $Vol_L$, and magnetic properties $\mu_i$, $B_{sat}$) can be obtained from Lookup Table I.

In addition, to build the electrical simulation, parasitic parameters $R_L$ of inductor, and $ESR$ and $ESL$ of capacitor are required. With Lookup Table I for inductors, the equivalent resistance of inductor $R_L$ can be computed with (3), where $r$ is the resistance of wire per unit length. With Lookup Table II for capacitors, the equivalent series resistance $ESR$ and the equivalent series inductance $ESL$ of capacitor can be computed with (4) and (5) [34], respectively, where $\tan\delta$ and $k_{esl}$ are the dissipation factor and $ESL$ factor, both of which are obtained with the capacitor database.

$$R_L = N \cdot (OD - ID + 2H) \cdot r \quad (3)$$

$$ESR = \frac{\tan\delta}{2\pi f_s C} \quad (4)$$

$$ESL = k_{esl}/C \quad (5)$$

Magnetic simulation tool, Ansys, is used to evaluate magnetic core loss $P_{l\_L\_Fe}$. With necessary features of inductor core provided by Lookup Table I, magnetic simulation models can be built to evaluate $P_{l\_L\_Fe}$ of all the selected combinations of $f_s$, $L$, $C$.

Electrical simulation tool, LTspice, is utilized to evaluate electrical losses, voltage ripple and current ripple. With the Spice model of power switches provided by the manufacturer, LTspice can evaluate the losses of high-side switch $P_{l\_s1}$ and low-side switch $P_{l\_s2}$, in which the switching and conduction losses are included. Besides, with the essential parasitic parameters obtained, electrical simulation by LTspice can also evaluate the inductor copper loss $P_{l\_L\_Cu}$, capacitor loss $P_{l\_C}$. Moreover, voltage and current ripples $\Delta V_o\%$, $\Delta I_L\%$ of all the combinations of $f_s$, $L$, $C$ are also evaluated with LTspice to generate training data for BN-NN.

In this process, programming language, such as Python, can be adopted to automate the running of simulations through proper interfaces of the simulation tools. Consequently, the programming language can automatically adjust parameters, conduct running of simulations and collect performance data.

(*c*) *Step* 3: *Build Data-Driven Models of Power Losses and Ripples via BN-NN*

After implementation of simulations in Step 2 of Stage 3, the losses of switch, inductor losses, capacitor loss and voltage and current ripples of all the combinations of $f_s$, $L$, $C$ have been assessed. With these limited performance data, BN-NN will be trained so that the performance of any possible designs can be evaluated.

As introduced in Section II-B (a), BN-NN in Fig. 3 which is good at avoiding over-fitting problem is specially adopted.

The adopted BN-NN includes three types of layers: scalarization layer, hidden layers, and output layer.

In the AI-D approach for the parameter design of power converters, the first scalarization layer aims to rescale the magnitude of design parameters $f_s$, $L$, $C$ within the range of [0, 1]. For instance, originally, $f_s$ is more than tens of kilohertz, while $L$ is smaller than millihenry. They will be both rescaled into [0, 1] for unbiased BN-NN training.

There is assumed to exist $H$ hidden layers, and each hidden layer includes $N_h$ neurons. Every hidden layer contains a fully connected layer followed by a batch-normalization layer. The batch-normalization layer aims to avoid the over-fitting problem, which can seriously deteriorate the NN accuracy.

The last layer is the output layer, and the outputs include losses $P_{l\_s1}$, $P_{l\_s2}$, $P_{l\_L\_Cu}$, $P_{l\_L\_Fe}$, $P_{l\_C}$ and ripples $\Delta V_o\%$, $\Delta I_L\%$. The complete structure of BN-NN is given in Fig. 10.

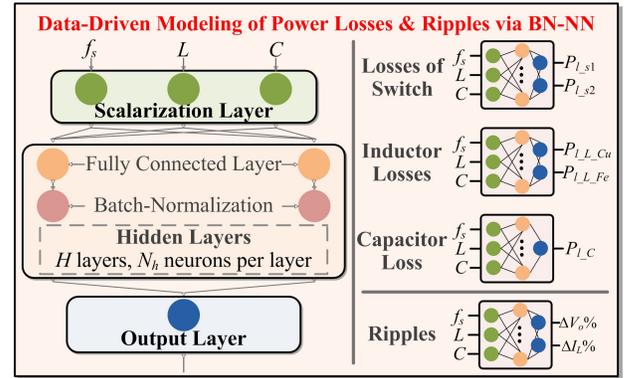

Fig. 10. Data-driven modeling of power losses and ripples via BN-NN.

The performance data of power losses and ripples which are offered by simulations of all the selected $N_1 \times N_2 \times N_3$ combinations are divided into training (70%), validating (15%) and testing (15%) sets, which are used to train BN-NN, to select structure of BN-NN ($N_h$, $H$), and to test BN-NN in new unseen designs, respectively. With the training and validating sets, different options of $N_h$ and $H$ are tried, among them the particular $N_h$ and $H$ that reach the least error on the validating set are the selected BN-NN structure.

Until here, the automatic analysis and deduction process has been finished.

*D. Stage* 4: *Search for Optimal Design Parameters $f_s^*$, $L^*$, $C^*$ via Genetic Algorithm*

To achieve optimal parameter design for power converters, GA is utilized in this stage to find optimal design parameters $f_s$, $L$, $C$ with design constraints in volume, and current and voltage ripples. The optimization problem is expressed with (6).

$$\min_{f_s, L, C} \left( P_{l\_s1} + P_{l\_s2} + P_{l\_L\_Cu} + P_{l\_L\_Fe} + P_{l\_C} \right)$$
$$\text{subject to:} \quad Vol_L + Vol_C \leq Vol_{lim}, \quad (6)$$
$$\Delta V_o\% \leq (\Delta V_o\%)_{lim},$$
$$\Delta I_L\% \leq (\Delta I_L\%)_{lim}.$$

The detailed flowchart of GA in solving (6) and searching for the globally optimal $f_s^*$, $L^*$, $C^*$ is shown in Fig. 11, and briefly illustrated as follows. The fitness value of the $i^{th}$ individual $F_i$ is computed with (7a), where $O_i$ is the $i^{th}$ objective value as expressed in (7b), $O_{max}$ and $O_{min}$ are the maximum and minimum of $O_i$ for all individuals, and $\xi$ is a small constant. In (7b), $P_{l\_tot}$ is the total power loss, which is shown in (7c).

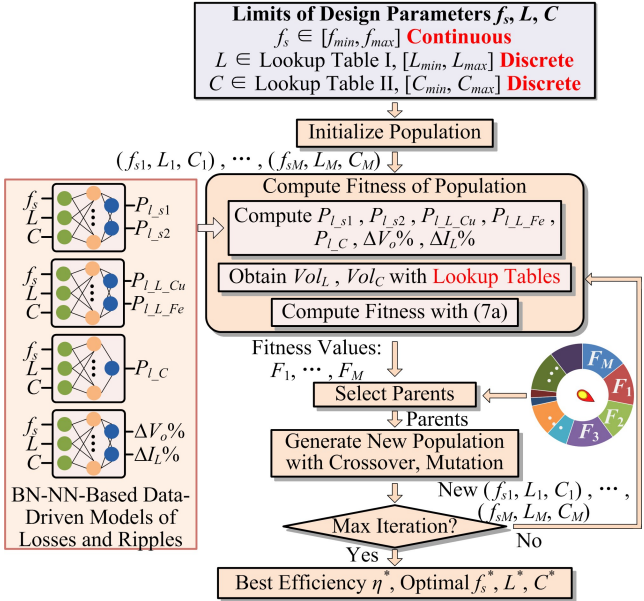

Fig. 11. Flowchart of GA in searching for the globally optimal $f_s^*$, $L^*$, $C^*$.

$$F_i = \frac{O_{\max} - O_i}{O_{\max} - O_{\min}} + \xi \quad (7a)$$

$$O_i = P_{l\_tot} + \max\left(0, \frac{Vol}{Vol_{lim}} - 1\right) \\ + \max\left(0, \frac{\Delta V_o\%}{(\Delta V_o\%)_{lim}} - 1\right) + \max\left(0, \frac{\Delta I_L\%}{(\Delta I_L\%)_{lim}} - 1\right) \quad (7b)$$

$$P_{l\_tot} = P_{l\_s1} + P_{l\_s2} + P_{l\_L\_Cu} + P_{l\_L\_Fe} + P_{l\_C} \quad (7c)$$

To restrict the volume and ripples of the designed converters, the objective value $O_i$ in (7b) has introduced penalty terms such as "$\max(0, Vol/Vol_{lim} - 1)$", so that the negative effects of these constraints being exceeded are considered.

Till now, the optimization process has been finished with the assistance of GA. Thus, all stages of the proposed AI-based design approach have completed, and the optimal synchronous Buck converter with satisfactory comprehensive performance has been designed.

## IV. DESIGN CASE OF THE PROPOSED AI-D APPROACH TO DESIGN AN EFFICIENCY-OPTIMAL SYNCHRONOUS BUCK CONVERTER IN EV

With the proposed AI-D approach elaborated in Section III, an efficiency-optimal synchronous Buck converter applied in 48 V to 12 V accessory-load power supply system of EV [31] is designed. The design case is illustrated below stage by stage.

### A. Determine Design Specifications

In Stage 1, design conditions and requirements of the synchronous Buck converter applied in 48 V to 12 V accessory-load power supply system of EV are specified, as listed in Table I. Rated power $P_o$ is selected as 100 W for the accessory loads in the EV power supply system [32], [35].

TABLE I. DESIGN SPECIFICATIONS

| Operating Specifications | | | |
|---|---|---|---|
| Topology | Synchronous Buck | $V_o$ | 12 V |
| $V_{in}$ | 48 V | $P_o$ | 100 W |
| Power Switches | | | |
| Switches | IRFB4310PbF, Infineon | Dead time | 200 ns |
| $R_{DS(on)}$ | 5.6 m$\Omega$ | $V_{DSS}$ | 100 V |
| Output LC Filter | | | |
| Inductor cores | | Toroidal TAF-200 series | |
| Inductor wire | | UEFN/U 1mm | |
| Capacitors | | 25 V Nippon KZE series | |
| Design Parameters | | | |
| Switching frequency $f_s$ | Inductance $L$ | | Capacitance $C$ |
| Limits of Design Parameters | | | |
| $f_s$ | | $f_{min}$ = 20 kHz; $f_{max}$ = 200 kHz | |
| $L$ | | $L_{min}$ = 30 $\mu$H; $L_{max}$ = 2 mH | |
| $C$ | | $C_{min}$ = 20 $\mu$F; $C_{max}$ = 1000 $\mu$F | |
| Design Constraints | | | |
| Volume $Vol$ | | $\leq Vol_{lim}$ = 7 cm$^3$ | |
| Voltage ripple $\Delta V_o\%$ | | $\leq (\Delta V_o\%)_{lim}$ = 1% | |
| Current ripple $\Delta I_L\%$ | | $\leq (\Delta I_L\%)_{lim}$ = 10% | |

When the limits of design parameters are determined, switching frequency $f_s$ is selected within the suitable range [20 kHz, 200 kHz] according to [36], [37]. As determined by the range of $f_s$, the ranges of $L$ and $C$ are chosen from several tens of $\mu$H, $\mu$F to mH, mF, where the selection of upper limits $L_{max}$ and $C_{max}$ is out of cost and power density perspectives [25], and the selection of lower limits $L_{min}$ and $C_{min}$ considers filtering performance [8]. If a specific application requires different ranges of $f_s$, $L$ and $C$ from the given ranges in this design case, the proposed AI-D approach can still be applied with no changes in any steps.

In terms of power switch selection, Infineon IRFB4310PbF is chosen with the considerations of smaller drain-source on-resistance and suitable drain-source breakdown voltage under the given operating specifications. The proposed AI-D approach is still applicable if other power switches are considered. It should be noticed that the simulations built in Stage 3 need to adopt the models of the used power switches.

### B. Create Lookup Tables for Inductors and Capacitors

In Stage 2, lookup tables for inductors and capacitors are created. In this design case, four toroidal cores (T80-75-200, T106-75-200, T131-75-200, T150-75-200) are adopted from TAF-200 series. By following the flowchart in Fig. 7, Lookup Table I for inductors is created, as shown in Fig. 12 and Fig. 13, in which the fill factor $K_u$ of inductor core is kept below 0.35 for easy manufacture [38]. Fig. 12 provides the information about the practical features of the selected cores, and Fig. 13 shows the possible discrete values of $L$ and corresponding core selections.

| Lookup Table I for Inductors: Features of Selected Cores | | | |
|---|---|---|---|
| T80-75-200 | T106-75-200 | T131-75-200 | T150-75-200 |
| OD 20.2 mm | OD 26.9 mm | OD 33 mm | OD 38.4 mm |
| ID 12.6 mm | ID 14.5 mm | ID 16.3 mm | ID 21.5 mm |
| H 6.35 mm | H 11.1 mm | H 11.1 mm | H 11.1 mm |
| $A_e$ 24.1 mm$^2$ | $A_e$ 68.8 mm$^2$ | $A_e$ 92.7 mm$^2$ | $A_e$ 93.8 mm$^2$ |
| $Vol_L$ 2.04 cm$^3$ | $Vol_L$ 6.31 cm$^3$ | $Vol_L$ 9.49 cm$^3$ | $Vol_L$ 12.9 cm$^3$ |
| $A_L$ 46 | $A_L$ 93 | $A_L$ 116 | $A_L$ 96 |
| $N_{max\_core}$ 56 | $N_{max\_core}$ 74 | $N_{max\_core}$ 93 | $N_{max\_core}$ 162 |
| $L_{max\_core}$ 144$\mu$H | $L_{max\_core}$ 509$\mu$H | $L_{max\_core}$ 1003$\mu$H | $L_{max\_core}$ 2519$\mu$H |

$L_{min}$ — 144$\mu$H — 509$\mu$H — 1003$\mu$H — $L_{max}$

/* 144$\mu$H, 509$\mu$H, ... are the reachable maximum inductance of each core $L_{max\_core}$, as computed by 0.35 fill factor */

Fig. 12. Lookup Table I for inductors: features of the selected cores.

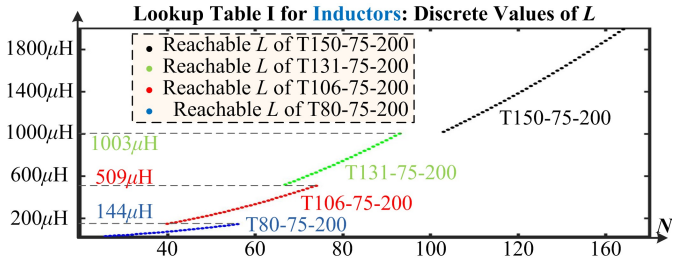

Fig. 13. Lookup Table I for inductors: discrete values of $L$.

| Lookup Table II for **Capacitors**: Features of Selected Capacitors | | | | | | |
|---|---|---|---|---|---|---|
| Dissipation factor tan$\delta$ : 0.14 | | | ESL factor $k_{esl}$ : 4.389×10$^{-11}$ | | | |
| | $C_1$ | $C_2$ | $C_3$ | $C_4$ | $C_5$ | $C_6$ |
| **Capacitors** | 27$\mu$F | 47$\mu$F | 56$\mu$F | 100$\mu$F | 220$\mu$F | 330$\mu$F ... |
| $Vol_C$ (cm$^3$) | 0.137 | 0.216 | 0.218 | 0.343 | 0.578 | 0.754 |

Fig. 14. Lookup Table II for capacitors: features of selected capacitors.

| Lookup Table II for **Capacitors**: Discrete Values of $C$ | | | | | | |
|---|---|---|---|---|---|---|
| Single | $C_1$ | $C_2$ | $C_3$ | $C_4$ | $C_5$ | $C_6$ $C_7$ ... |
| Parallel | $C_1+C_1$ | $C_2+C_2$ | $C_1+C_3$ | $C_2+C_3$ | $C_1+C_4$ | ... |
| Serial | $C_1//C_1$ | $C_2//C_2$ | $C_1//C_3$ | $C_2//C_3$ | $C_1//C_4$ | ... |
| Parallel and Serial | $(C_1+C_1)//C_1$ | | $(C_1+C_2)//C_3$ | | $(C_1+C_3)//C_2$ | ... |

Fig. 15. Lookup Table II for capacitors: discrete values of $C$.

Afterwards, with the maximum number of capacitors for parallel and serial connections $MP$ selected as 5, according to the flowchart in Fig. 8, Lookup Table II for capacitors is created as shown in Fig. 14 and Fig. 15. Fig. 14 shows the features of selected capacitors, and Fig. 15 describes the possible discrete values of $C$ considering parallel and serial connections.

### C. Build Data-Driven Models for Power Losses and Ripples

In Stage 3, data-driven models for power losses and ripples with BN-NN are built by following Fig. 9, and the detailed three steps of Stage 3 are illustrated as follows.

- In Step 1, within the range of $f_s$, $L$ and $C$ defined in Table I, $N_1$, $N_2$ and $N_3$ in Fig. 9 are set as 20, 20 and 20, respectively, and thus 20×20×20=8000 number of combinations of design parameters in total are generated.
- In Step 2, magnetic simulation model in Ansys and electrical simulation model in LTspice are built. Simulations are conducted to evaluate the power losses and ripples of all 8000 combinations, which serve as training data for BN-NN.
- In Step 3, four BN-NN are built: BN-NN for losses of switch ($P_{l\_s1}$, $P_{l\_s2}$) includes 3 hidden layers and 10 neurons per layer; BN-NN for inductor losses ($P_{l\_L\_Cu}$, $P_{l\_L\_Fe}$) has 3 hidden layers and 20 neurons per layer; BN-NN for capacitor loss $P_{l\_C}$ incorporates 2 hidden layers, each of which has 10 neurons; BN-NN for ripples ($\Delta V_o\%$, $\Delta I_L\%$) contains 2 hidden layers, and each layer has 10 neurons.

To reflect the higher generalization accuracy of BN-NN, several AI-based regression techniques (ridge regression, support vector regression, Bayesian regression, deep NN via Matlab Toolbox) are compared with, and the learning target is the power loss of the high-side switch $P_{l\_s1}$ as an example. As shown in Table II, among all the techniques compared, the proposed BN-NN in Fig. 10 manifests the smallest error on all the datasets, indicating the highest generalization accuracy.

TABLE II. MEAN-SQUARE-ERROR OF THE COMPARED REGRESSION METHODS

| | Training Set | Validating Set | Testing Set |
|---|---|---|---|
| Ridge regression | 0.719 | 0.689 | 0.712 |
| Support vector regression | 0.472 | 0.441 | 0.433 |
| Bayesian regression | 0.310 | 0.315 | 0.294 |
| NN via Matlab toolbox | 0.084 | 0.085 | 0.076 |
| **Proposed BN-NN** | **0.022** | **0.018** | **0.025** |

### D. Search for Optimal $f_s^*$, $L^*$, $C^*$ via GA

In Stage 4, globally optimal $f_s^*$, $L^*$, $C^*$ are found with GA in Fig. 11. With the trained data-driven models with BN-NN for power losses and ripples and Lookup Table I and II for volume, the optimal $f_s^*$, $L^*$, $C^*$ are searched. $f_s^*$ is searched within the continuous space $[f_{min}, f_{max}]$, $L$ is chosen from Lookup Table I which lies within $[L_{min}, L_{max}]$, and discrete $C$ is chosen from Lookup Table II which lies within $[C_{min}, C_{max}]$.

The finalized optimal design results of $f_s^*$, $L^*$ and $C^*$ are 36.6 $k$Hz, 281.3 $\mu$H and 112 $\mu$F, respectively. They provide an optimal efficiency $\eta^*$ at 93.85%,

The reason why $f_s^*$ is optimized to 36.6 $k$Hz is related with the predetermined volume constraint $Vol_{lim}$. If high power density is expected and $Vol_{lim}$ is set at a smaller value compared with the one set in this design case, the optimized $f_s^*$ may reach a bigger value. But in this situation, the efficiency will decrease due to the increase of switching and core losses.

TABLE III. DESIGNED CONVERTER IN THE 48 V TO 12 V ACCESSORY-LOAD POWER SUPPLY SYSTEM OF EV VIA THE PROPOSED AI-D APPROACH

| Designed Efficiency-Optimal Synchronous Buck Converter in EV | |
|---|---|
| Topology | Synchronous Buck |
| Switch | IRFB4310PbF, Infineon |
| $f_s$ | 36.6 $k$Hz |
| $L$ | 281.3$\mu$H, core T106-75-200, wire UEFN/U 1mm, number of turns 55 |
| $C$ | 112 $\mu$F, 2 number of 56 $\mu$F of 25 V Nippon KZE in parallel |

TABLE IV. THEORETICAL PERFORMANCE OF THE DESIGNED SYNCHRONOUS BUCK CONVERTER

| Theoretical Performance of Designed Converter | |
|---|---|
| Efficiency $\eta$ | 93.85% |
| Volume $Vol_L+Vol_C$ | 6.746 cm$^3$ |
| Ripples $\Delta V_o\%$ and $\Delta I_L\%$ | $\Delta V_o\%$ = 0.573%; $\Delta I_L\%$ = 9.7% |

The design results for the efficiency-optimal synchronous Buck converter in the 48 V to 12 V accessory-load power supply system of EV via the proposed AI-D approach are summarized in Table III. The theoretical performance of the designed converter is listed in Table IV.

### E. Average CPU Execution Time for Applying the Proposed AI-D Approach in the Design Case

TABLE V. AVERAGE CPU EXECUTION TIME OF AI-D APPROACH

| Stages in AI-D | Average CPU Execution Time |
|---|---|
| Stage 2 | Total 0.232 seconds |
| Stage 3: Step 1 | Total 0.074 seconds |
| **Stage 3: Step 2** | **Total 2 days and 4 hours to run the required simulations using four CPU cores** |
| Stage 3: Step 3 | 1 minute 13.4 seconds to train all NNs |
| Stage 4 | Total 24.38 seconds |

To provide insights of the computational time required to implement the proposed AI-D approach, under the computer platform with Intel Xeon CPU E5-1630 @ 3.7 GHz, 16 GB RAM, and Windows 10 operating system, the average CPU execution time of the proposed AI-D approach applied in the design case is given in Table V. Based on Table V, most of the computational time and resources are spent on running simulations to collect power loss and ripple performance data, while the CPU execution time of other stages is neglectable.

## V. EXPERIMENTAL VERIFICATION

In this section, to further verify the designed synchronous Buck converter in Table III for the 48 V to 12 V accessory-load power supply system of EV, a prototype has been built and hardware experiments have been conducted. The hardware platform is shown in Fig. 16.

### A. Steady-State Waveforms of the Designed Optimally Efficient Synchronous Buck Converter

Under the rated conditions given by Table I, the waveforms of the designed converter in steady state are shown in Fig. 17, where the notation and direction of waveforms are indicated in Fig. 4.

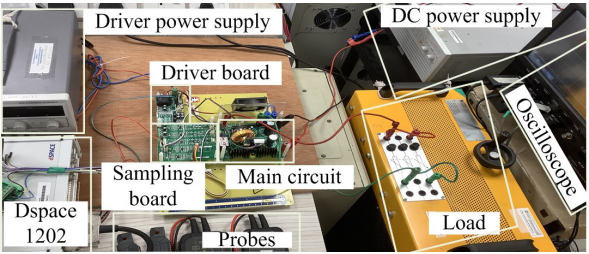

Fig. 16. The hardware platform of the designed DC-DC converter.

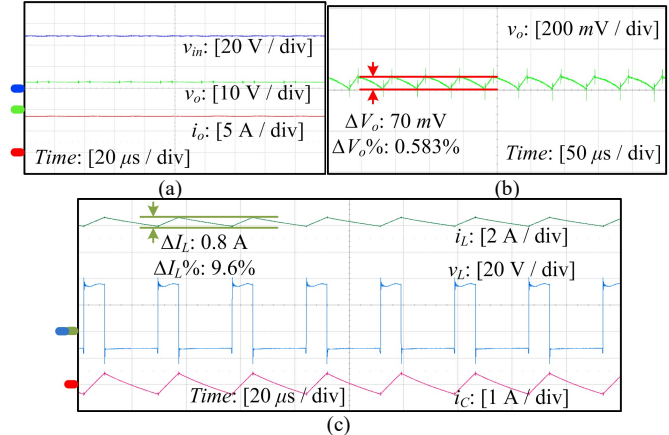

Fig. 17. Steady-state waveforms of the designed efficiency-optimal converter: (a) $v_{in}$, $v_o$, $i_o$; (b) zoom-in view of $v_o$; (c) $v_L$, $i_L$, $i_C$.

### B. Experimental Efficiency of the Designed Converter

(*a*) *Validation of the Optimal Efficiency of the Designed Synchronous Buck Converter when $f_s$, $L$, $C$ are Varying*

To verify that the designed converter reaches an optimal efficiency, the following experiments are implemented. The switching frequency $f_s$ varies around the optimal frequency $f_s^*$, as shown in Fig. 18 (a), where the efficiency reaches the highest at the designed optimal $f_s^*$. In addition, when $L$ varies around the optimal $L^*$, as shown in Fig. 18 (b), the values smaller than $L^*$ should be avoided because they fail to meet the required 10% current ripple constraint in equation (6), even though they achieve higher efficiency. Among the $L$ values no smaller than $L^*$, the selected value $L^*$ displays best efficiency, so the optimal efficiency under given constraints is still achieved at the designed optimal $L^*$. Besides, as shown in Fig. 18 (c), $C$ varies within [81$\mu$F, 168$\mu$F], and the results verify that the optimal efficiency at the designed $C^*$ is achieved. Consequently, based on Fig. 18 (a) to (c), the designed synchronous Buck converter applied in EV enjoys the optimal efficiency.

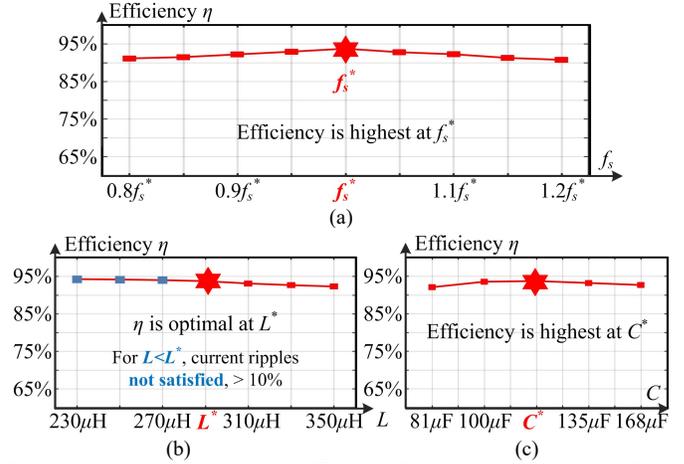

Fig. 18. Validation of the optimal efficiency of the designed synchronous Buck converter: (a) $f_s$ varies within [0.8$f_s^*$, 1.2$f_s^*$]; (b) $L$ varies within [230$\mu$H, 350$\mu$H]; (c) $C$ varies within [81$\mu$F, 168$\mu$F].

(*b*) *Efficiency Comparison of Designed Synchronous Buck Converters via the Proposed AI-D Approach, the CAO Approach and the Conventional Approach*

To validate the superiority of the proposed AI-D approach, the conventional approach [34] and a CAO approach [8] are compared with it. Conventionally, voltage and current ripple constraints are used to determine the values of $L$ and $C$ [34], which are computed with (8), where $V_o$, $I_o$, $f_s$, $(\Delta V_o\%)_{lim}$ and $(\Delta I_L\%)_{lim}$ are given in Table I. And the computed $L$ and $C$ are listed in Table VI. As detailly discussed in [8], the compared CAO approach manually analyzes and deduces the expressions of total power loss, and then optimizes the power loss expressions for optimally efficient synchronous Buck converter. The designed $L$ and $C$ via the compared CAO are shown in Table VII.

$$L = \frac{(1-D) \cdot V_o}{f_s \cdot (\Delta I_L\%)_{lim} \cdot I_o} \quad (8a)$$

$$C = \frac{(\pi + 4\tan\delta) \cdot (\Delta I_L\%)_{lim} \cdot I_o}{8\pi f_s \cdot V_o \cdot (\Delta V_o\%)_{lim}} \quad (8b)$$

TABLE VI. CONVENTIONALLY DESIGNED SYNCHRONOUS BUCK CONVERTER

| | Conventionally Designed Synchronous Buck Converter |
|---|---|
| $f_s$ | 20 *k*Hz |
| $L$ | 540$\mu$H, core T135-75-200, wire UEFN/U 1mm, number of turns 68 |
| $C$ | 56 $\mu$F, single 56 $\mu$F of 25 V Nippon KZE |

TABLE VII. CAO-DESIGNED SYNCHRONOUS BUCK CONVERTER

| | CAO-Designed Synchronous Buck Converter |
|---|---|
| $f_s$ | 36.6 *k*Hz |
| $L$ | 334.8$\mu$H, core T106-75-200, wire UEFN/U 1mm, number of turns 60 |
| $C$ | 94 $\mu$F, 2 number of 47 $\mu$F of 25 V Nippon KZE in parallel |

The comparison of experimental efficiency is shown in Fig. 19. Obviously, the designed converter via the proposed AI-D approach achieves the highest efficiency compared with the conventionally designed one and the one with CAO approach under different load levels. The peak efficiency of the AI-D designed converter in Table III reaches 93.68%, while the peak efficiency of the conventionally designed converter in Table VI is only 90.16%. Compared with the 92.46% achieved by the design via CAO in Table VII, the efficiency of the design via the proposed AI-D is 1.22% higher.

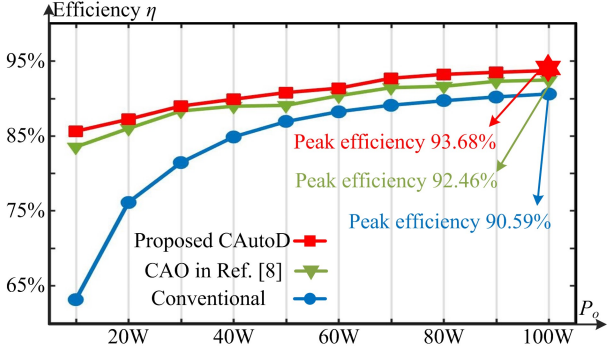

Fig. 19. Experimental efficiency comparison of designed converters via the proposed AI-D approach, the CAO approach and the conventional approach.

### C. Experimental Volume and Ripples of the Designed Converter

In terms of the volume of the designed converter via AI-D approach, the volume of inductor and capacitor is measured as 6.9 cm$^3$, which meets the volume constraint of 7 cm$^3$.

As for the experimental ripples, based on Fig. 17, the experimental voltage ripple $\Delta V_o$% and current ripple $\Delta I_L$% are 0.583% and 9.6%, respectively, both of which meet the corresponding constraints at 1% and 10%.

### D. Comparison between the Experimental and Theoretical Efficiency, Volume and Ripples of the Designed Converter

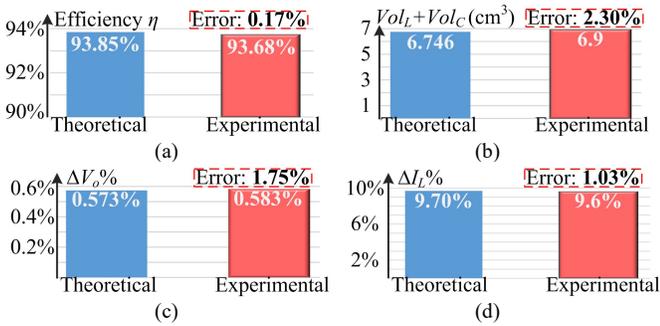

Fig. 20. Comparison between the experimental and theoretical performance: (a) efficiency $\eta$; (b) volume $Vol_C+Vol_L$; (c) voltage ripple $\Delta V_o$%; (d) current ripple $\Delta I_L$%.

In this part, the experimental power losses, volume and ripples are compared with the theoretical performance in Table IV. As shown in Fig. 20, the experimental performance is almost the same as those in theory, and the average error is only 1.31%. This proves the feasibility and high accuracy of the proposed AI-D approach thanks to the specially adopted BN-NN in analysis and deduction process and the adopted GA in optimization process.

## VI. CONCLUSION

An artificial-intelligence-based design (AI-D) approach is proposed in this paper for the parameter design of power converters. This proposed AI-D approach is able to conduct the analysis and deduction process and the optimization process in an automatic fashion. It provides two outstanding advantages: Firstly, it greatly relieves the work burden for engineers and realizes a fast and easy-implemented design process; Secondly, high design accuracy can also be ensured because of the mitigated human-dependence.

In the proposed AI-D approach, to achieve automation in the analysis and deduction process, simulation tools and batch-normalization neural network (BN-NN) are adopted to build data-driven models for the optimization objectives and design constraints. The specially utilized BN-NN is beneficial for design accuracy because it is good at avoiding over-fitting problem. Besides, to achieve automation in the optimization process, genetic algorithm is used to search for optimal design results.

The proposed AI-D approach is validated in the parameter design of an efficiency-oriented synchronous Buck converter in the 48 V to 12 V accessory-load supply system in EV. And hardware experiments have validated the feasibility and high accuracy of the proposed AI-D approach.

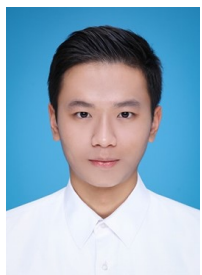

**Xinze Li** received his bachelor degree in Electrical Engineering and its Automation from Shandong University, China, 2018. He is currently a full-time Ph.D. candidate with the School of Electrical and Electronic Engineering in Nanyang Technological University, Singapore.

His research interests include parameter design of DC-DC converter, and applications of evolutionary algorithms and deep learning algorithms in power electronics.

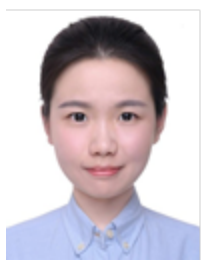

**Fanfan Lin** was born in Fujian, China in 1996. She received her bachelor degree in electrical engineering from Harbin Institute of Technology in China in 2018. From 2018, she studies as a Ph. D. student in Nanyang Technological University in Singapore. Her research interest includes the power converter design with artificial intelligence.

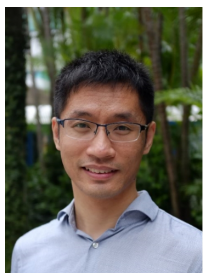

**Xin Zhang** (M'15-SM'20) received the Ph.D. degree in Automatic Control and Systems Engineering from the University of Sheffield, U.K., in 2016 and the Ph.D. degree in Electronic and Electrical Engineering from Nanjing University of Aeronautics & Astronautics, China, in 2014.

From February 2017 to December 2020, he was an Assistant Professor of power engineering with the School of Electrical and Electronic Engineering, Nanyang Technological University, Singapore. Currently, he is the professor at Zhejiang University. He is generally interested in power electronics, power systems, and advanced control theory, together with their applications in various sectors.

Dr. Zhang has received the Highly Prestigious Chinese National Award for Outstanding Students Abroad in 2016.

He is the Associated Editor of IEEE TIE/JESTPE/OJPE/Access and IET Power Electronics.

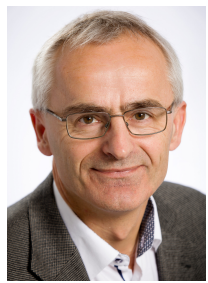

**Frede Blaabjerg** (S'86–M'88–SM'97–F'03) received the Ph.D. degree in electrical engineering from Aalborg University, Aalborg, Denmark, in 1995, and the honoris causa degree from the Univer-sity Politehnica Timisoara the University Politehnica Timisoara (UPT), Timisoara, Romania, in 2017, and Tallinn Technical University (TTU), Tallinn, Estonia in 2018.

He was with ABB-Scandia, Randers, Denmark, from 1987 to 1988. He became an Assistant Professor with the Department of Energy Technology, Aalborg University, in 1992, an Associate Professor in 1996, and a Full Professor of power electronics and drives in 1998, where he has been a Villum Investigator since 2017. He has authored or coauthored more than 600 journal articles in different fields of power electronics and its applications, coauthored four monographs, and edited ten books in power electronics and its applications. His current research interests include power electronics and its applications, such as in wind turbines, photovoltaic systems, reliability, harmonics, and adjustable speed drives.

Dr. Blaabjerg received the 31 IEEE Prize Paper Awards, the IEEE PELS Distinguished Service Award in 2009, the EPE-PEMC Council Award in 2010, the IEEE William E. Newell Power Electronics Award 2014, and the Villum Kann Rasmussen Research Award 2014. He was a recipient of the Global Energy Award for a significant contribution to the development of technologies that provide new energy development opportunities in 2019. He was a recipient of the IEEE Edison Medal in 2020. He is the President of the IEEE Power Electronics Society for 2019–2020 and the Vice President of the Danish Academy of Technical Sciences. He was the Editor-in-Chief of the IEEE IEEE TRANSACTIONS ON POWER ELECTRONICS FROM 2006 to 2012. He was a Distinguished Lecturer of the IEEE Power Electronics Society from 2005 to 2007 and the IEEE Industry Applications Society from 2010 to 2011 and 2017 to 2018. He was nominated by Thomson Reuters to be included in the 250 most cited researchers in engineering in the world in 2014–2018.